\input{aipcheck}

\documentclass[
    ,final            
  ]
  {aipproc}

\layoutstyle{6x9}

\usepackage{epstopdf}
\begin{document}

\title{A  QCD motivated model for soft processes}


\author{A.~Kormilitzin}{
address={Department of Particle Physics, School of Physics and Astronomy
Raymond and Beverly Sackler
 Faculty
of Exact Science , Tel Aviv University, Tel Aviv, 69978, Israel
}
}

\author{E.~Levin}{
,altaddress=
  {Department of Particle Physics, School of Physics and Astronomy
Raymond and Beverly Sackler
 Faculty
of Exact Science , Tel Aviv University, Tel Aviv, 69978, Israel
}
}

\begin{abstract}
In this talk we give a brief description of a QCD motivated model for  both hard and soft interactions at high energies. In this model
the long distance behaviour of the scattering amplitude  is determined  by the dipole scattering amplitude in the saturation domain. All phenomenological parameters for dipole-proton interaction were fitted from the deep inelastic scattering data and the soft processes are described with only one new parameter, related to the wave function of hadron. It turns out that we do not need  to introduce the so called soft Pomeron that has been used in high energy phenomenology for four decades.

\end{abstract}

\maketitle


\section{Main ideas and theoretical input}
This talk is the brief description of our attempt to find a self-consistent theoretical approach to soft (long distance) interaction
at high energy (see our paper of Ref.\cite{KOLE} for full presentation). Our main idea is  that there is no other dimemsionful  scales except the saturation scale for high energy interaction in the entire kinematic region. This idea looks strange since the traditional approach to soft high energy interaction is based on the phenomenological soft Pomeron, whose typical dimensionful scale is given by the slope of the Pomeron trajectory ($\alpha_P \,\approx 0.25 \,GeV^{-2}$). In simple words, we would like to replace the soft Pomeron exchange in the scattering amplitude by the QCD scattering amplitude in the saturation region.  $\alpha_P = 0$ for this amplitude and the saturation momentum ( $Q_s$) is the only dimensionful scale for it. Since this amplitude governs both short distances and long distances processes, all needed parameters in a such amplitude can be found from the DIS on the contrary to soft phenomenology in which all parameters are extracted from the soft scattering. The idea is not new and it is scattered in a number of reviews and talks (see references in Ref.\cite{KOLE}). However the first practical realization was suggested in Ref.\cite{BATAV} with an encouraging result: it is possible to describe the data on total cross section in the framework of such ideas.
In this talk as well as in our paper \cite{KOLE} we wish to check these ideas against the full set of experimental data in the region of accessible energies and even give some predictions for the LHC energy range.
Our approach is based on two  theoretical inputs.

{\bf Factorization of short and long distances:} Using the dispersion relation we can prove (see Ref.\cite{KOLE} that at least at large impact parameters the scattering amplitude for the BFKL Pomeron can be written in the form
\begin{equation} \label{SLF}
A_{BFKL} ( Q^2,x;b)\, =\, \stackrel{\mbox{ \small short distances}}{A_{BFKL}( Q^2,x;t=0)}\,\times\,{ \stackrel{\mbox{\small long  distances}}{ S(b)}}
\end{equation}
{\bf The scattering amplitude  of dipole (x,y) near to the saturation boundary } has the form\cite{LMP}:
\begin{equation} \label{AF}
N(Y-Y_0; x,y)\,=
\,1 - e^{- \int\,d^2 x' \,d^2 y'\,\, P( Y - Y_0;x,y; x',y') \gamma( Y_0; x',y')}
\end{equation}
where $\gamma( Y_0; x',y')$  is the scattering amplitude of dipole $( x',y')$ with the target at low energy
and $ P(Y - Y_0;x,y; x',y')$ is the  Green's function of the BFKL Pomeron.
\section{The model}
Using these two theory inputs we built our model assuming
\begin{itemize}
\item\,\,\,$\gamma_{mod}(Y_0; x',y') \, =\,
\,\,\frac{\pi^2\,\alpha_S(\mu^2_0)}{3}\,r^2\,x_0 G^{DGLAP}(x_0,\mu^2_0)\,S(b)$;
\item\,\,\,$ S(b)\,=\,\frac{2}{\pi \,R^2}(\frac{\sqrt{8}\,b}{R} )\,K_1 ( \frac{\sqrt{8}\,b}{R})$- the electromagnetic proton form factor;
\item\,\,\, Instead of the BFKL Pomeron in Eq.\ref{AF} we use the solution to the DGLAP equation;\
\item\,\,\,The scale of hardness in the DGLAP evolution ia chosen in the form:  $
\mu^2\, =\,\mu^2_0\,\,+\,\,\frac{C}{r^2}$.
\end{itemize}
Finally,
\begin{equation} \label{MO}
N(Y-Y_0; x,y)\,=\,1 - \exp\left(-\stackrel{\mbox{ \small short distances}}{\frac{\pi^2\,\alpha_S(\mu^2)}{3}\,r^2\, xG^{DGLAP}(x,\mu)}\,\times\,{ \stackrel{\mbox{ \small long distances}}{ S(b)}}\right)
\end{equation}
All parameters in Eq.\ref{MO} we found from fitting  the DIS data with $\chi^2$/d.o.f. =1.02.
The formulae for calculation of the physical observables are simple and can be found in Ref. \cite{KOLE}
and the structure of them is seen from the expression for the total cross section for meson proton scattering:  $ \sigma_{tot}\,=\,\int d^2b d^2 r |\Psi(r)|^2 \,N(Y - Y_0,r)$  where $\Psi$ is the wave function of a meson which consists of one colourless dipole
with size $r$, and $N$ is dipole scattering amplitude. Therefore, we need to specify the wave hadronic wave function in terms of t
dipoles. As have been mentioned, we assume that mesons consist of one dipole, for baryon we have two possibilities for the dipole content: two dipoles or three dipoles. We used both in our comparison with the experiment. For one dipole in the hadron we used the simplest Gaussian parameterization, namely $ \Psi( r ) \, = \frac{1}{\pi S} \,\exp( -\frac{ r^2_i}{S})$ where $S$ was chosen from comparison with the experiment. This is the only parameter that we found from the data on soft scattering. The second input that we need to calculate the scattering amplitude is to define the energy variable $x$ (or $Y-Y_0$).  We did this in the following way, using our the only dimensionful scale (saturation momentum $Q_s$):
\begin{equation}\label{x}
 x_{soft}\,=\,\frac{Q^2 + Q^2_s( x_{soft})}{s}\,\,\stackrel{Q^2 \to 0}{\rightarrow}\,\,\frac{Q^2_s( x_{soft})}{s}; \,\,\,\,
  x_{soft}\, =\,\frac{Q^2 + Q^2_s(x_{soft})}{s}\,\,\stackrel{Q^2 \gg Q^2_s}{\rightarrow}\,\,x_{Bjorken};
\end{equation}

\section{Comparison with the data}
Using the model we obtain the description of the experimental data which is not too bad. We consider that it is much better than we could hope with such a simple model.
\begin{figure}
\begin{tabular}{c c c}
\includegraphics[height=.2\textheight]{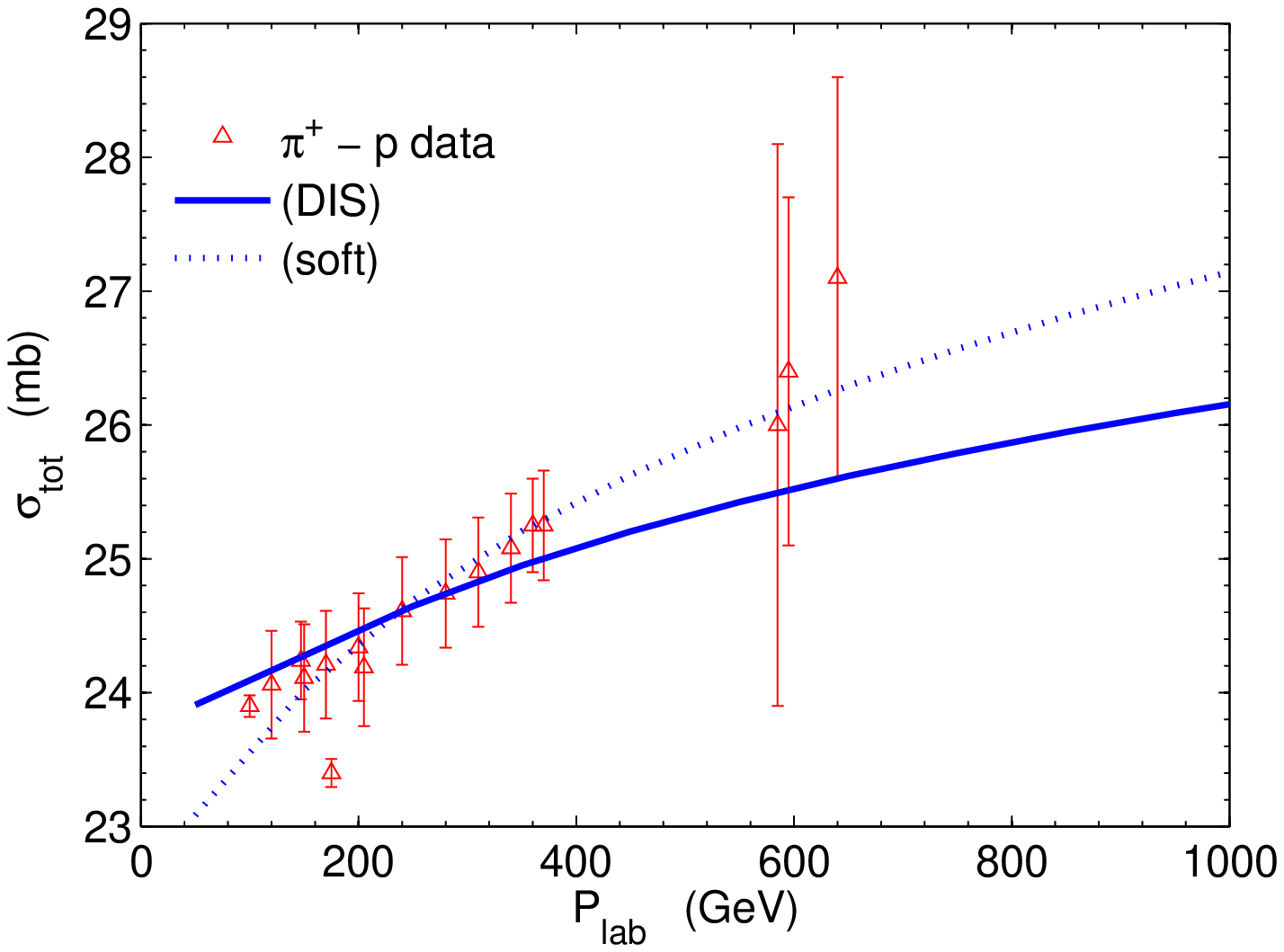}&
\includegraphics[height=.2\textheight]{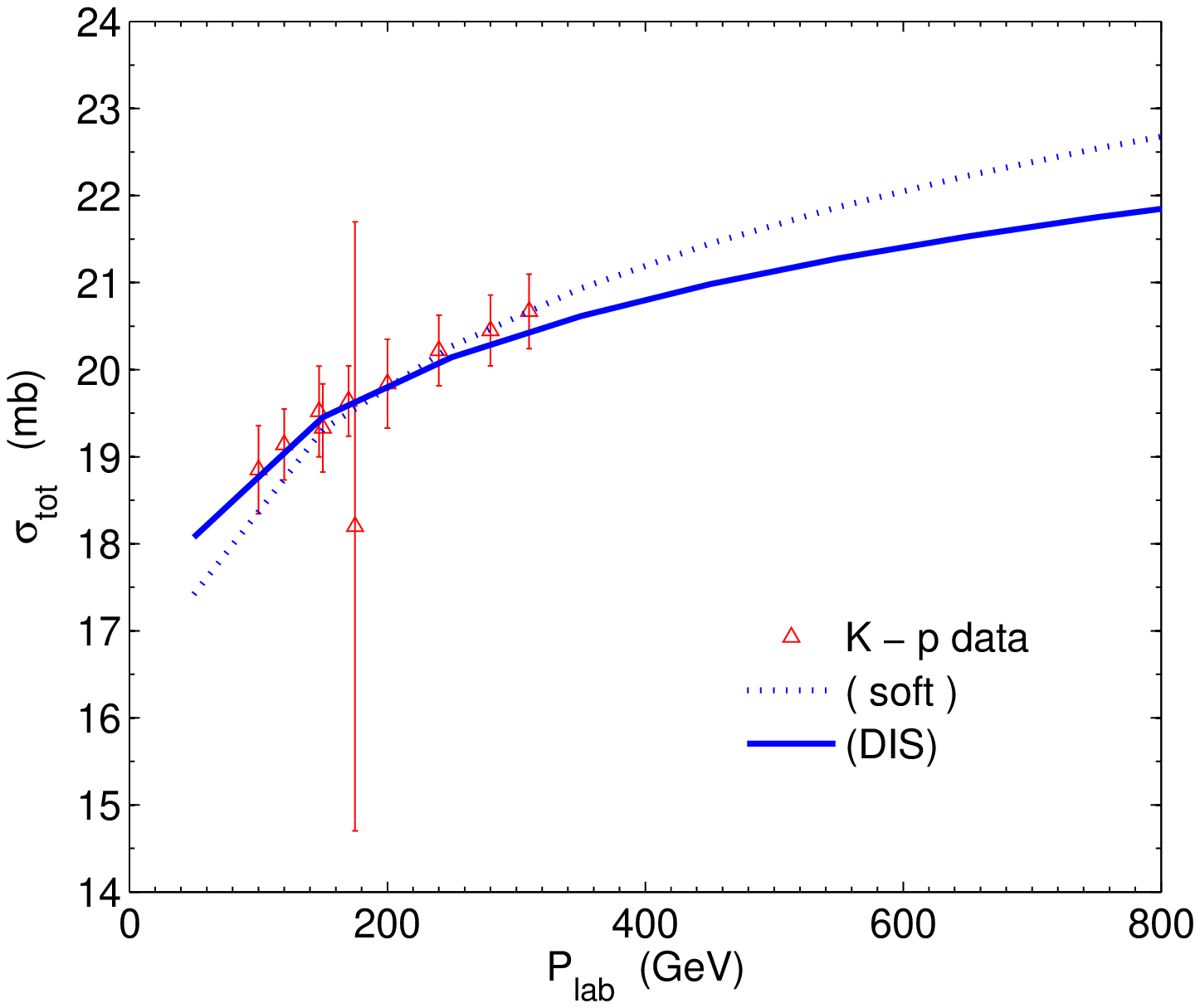} & 
\includegraphics[height=.2\textheight]{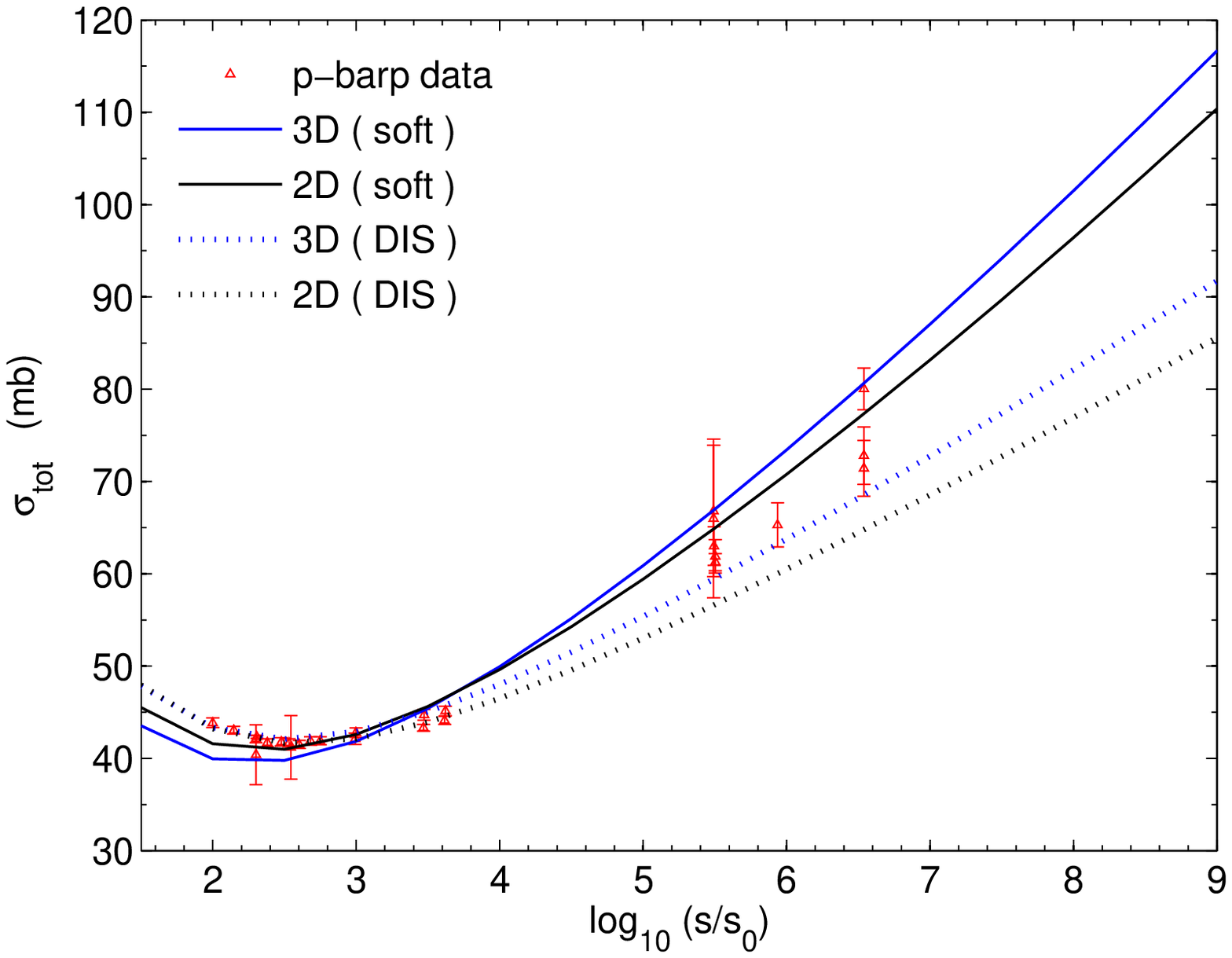}\\
  $\sigma_{tot}(\pi - p)$ & $\sigma_{tot}(K - p)$ &  $\sigma_{tot}(p  - \bar{p})$\\
\end{tabular}
\caption{}
\end{figure}

\begin{figure}
\begin{tabular}{c c}
\includegraphics[height=.2\textheight]{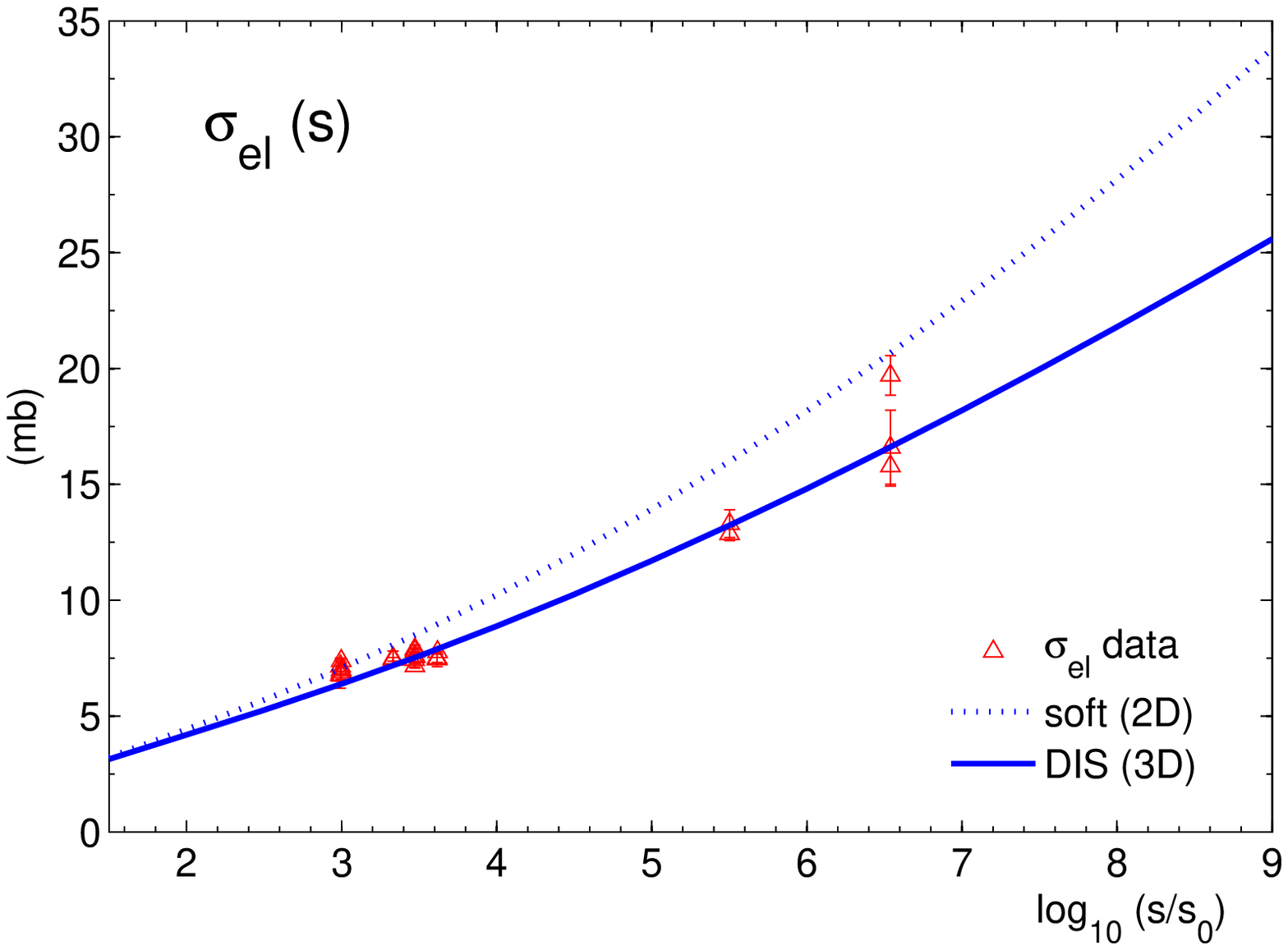}&
\includegraphics[height=.22\textheight]{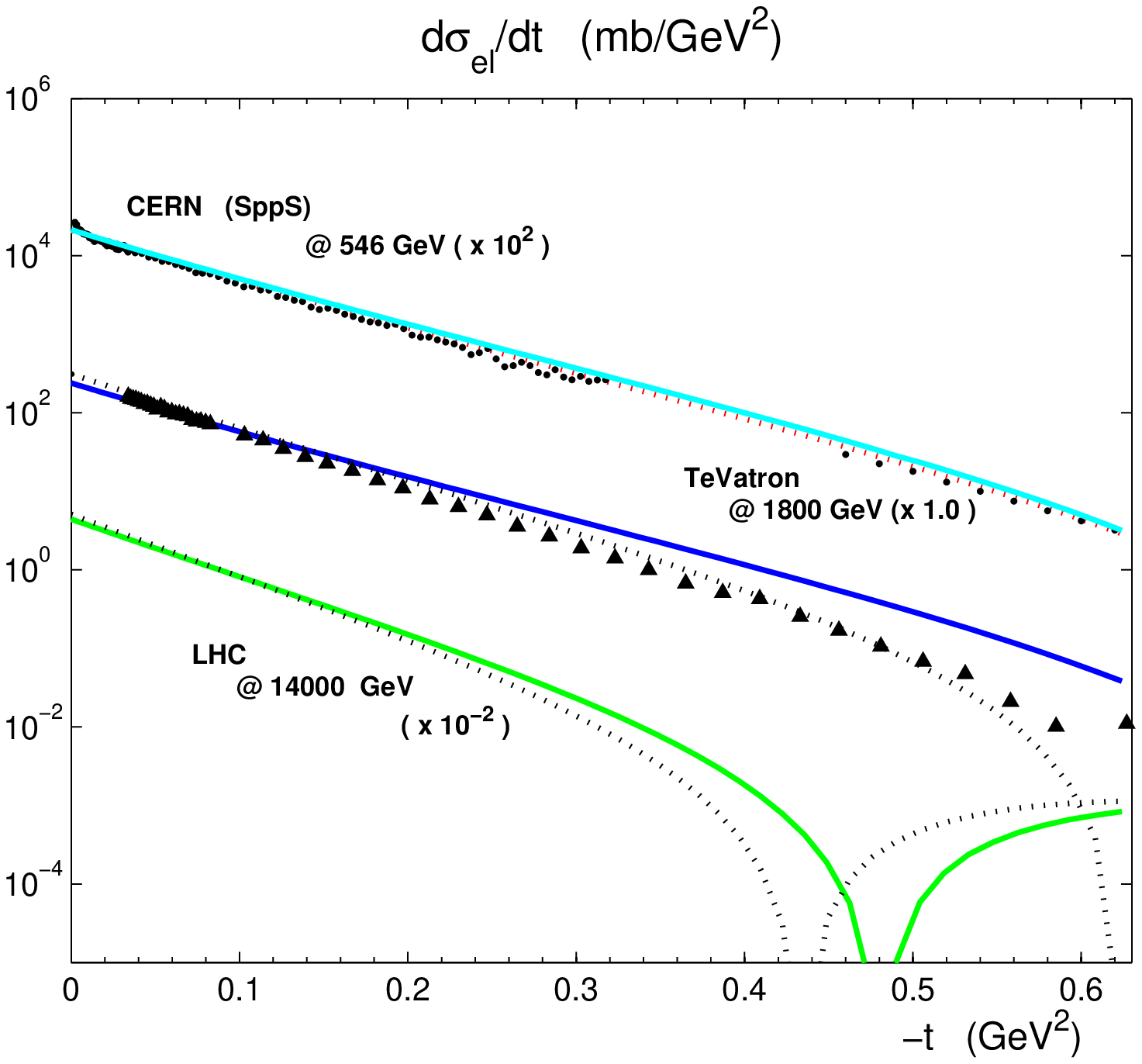}\\
  $\sigma_{el}(p  - \bar{p})$ & $d \sigma_{el}(p \bar{p})/d t $\\
\end{tabular}
\caption{}
\end{figure}

\begin{table}
\begin{tabular}{|l|l|l|l|l|l|}
\hline
          &               &                &         &        &       \\
Model &  $\sigma_{tot} $
&  $  \sigma_{el} $  & $ \sigma_{diff} $ & $B_{el}$ & $<|S^2|> $ \\
  & mb & mb & mb & $ GeV^{-2}$ & \\ \hline
Our A (DIS)  & 83.0 & 23.54   & 10 & 16.67 & (0.24-0.89)\%\\
Our B (Soft) & 101.3 & 28.84   & 10.5 & 18.4 &(0.24-0.57)\%  \\
GLM\cite{GLM1} & 110.5 & 25.3 & 11.6 & 20.5 & (0.7 -2)\% \\
GLMM\cite{GLM2} & 91.7 & 20.9& 11.8& 17.3 & 0.21\%\\
 RMK\cite{RMK}  & 88.0 (86.3)    & 20.1 (18.1)   & 13.3 (16.1)   & 19  & (1.2 - 3.2)\%\\
\hline
\end{tabular}
\caption{Predictions for the LHC}
\end{table}

One can see that our model reproduces quite well the energy and $t$ dependence of total and elastic cross sections. The most essential test for the model, however, is the dependence of the elastic slope versus energy. Indeed, in traditional high energy phenomenology this dependence was related to the value of $\alpha'_P\approx 0.25\,GeV^{-2}$ which in our model is equal to zero. Nevertheless, in spite of the fact that input $ \alpha'_P \,=\,0$ the shadowing corrections ,taken into account in our master formula of Eq.\ref{MO}, generate effective $\alpha'_p \approx 0.2\,GeV^{-2}$.  For the single diffraction cross section we take into account the emission of an additional gluon. The low curve in Fig. 3 corresponds to diffraction production of one dipole while the resulting curve includes the extra gluon emission. In all figures we plotted our calculations for two sets of parameterizations: set A (DIS) which gives very good description of DIS data with $\chi^2$/d.o.f. = 1.02; and set B(soft)    which  leads to worse description of DIS data ($\chi^2/d.o.f$- 3.6) but predicts the soft observables  closer to the experimental data. Notations 2D and 3D specify the model for nucleon with two or three dipoles.

\begin{figure}
\begin{tabular}{c c}
\includegraphics[height=.2\textheight]{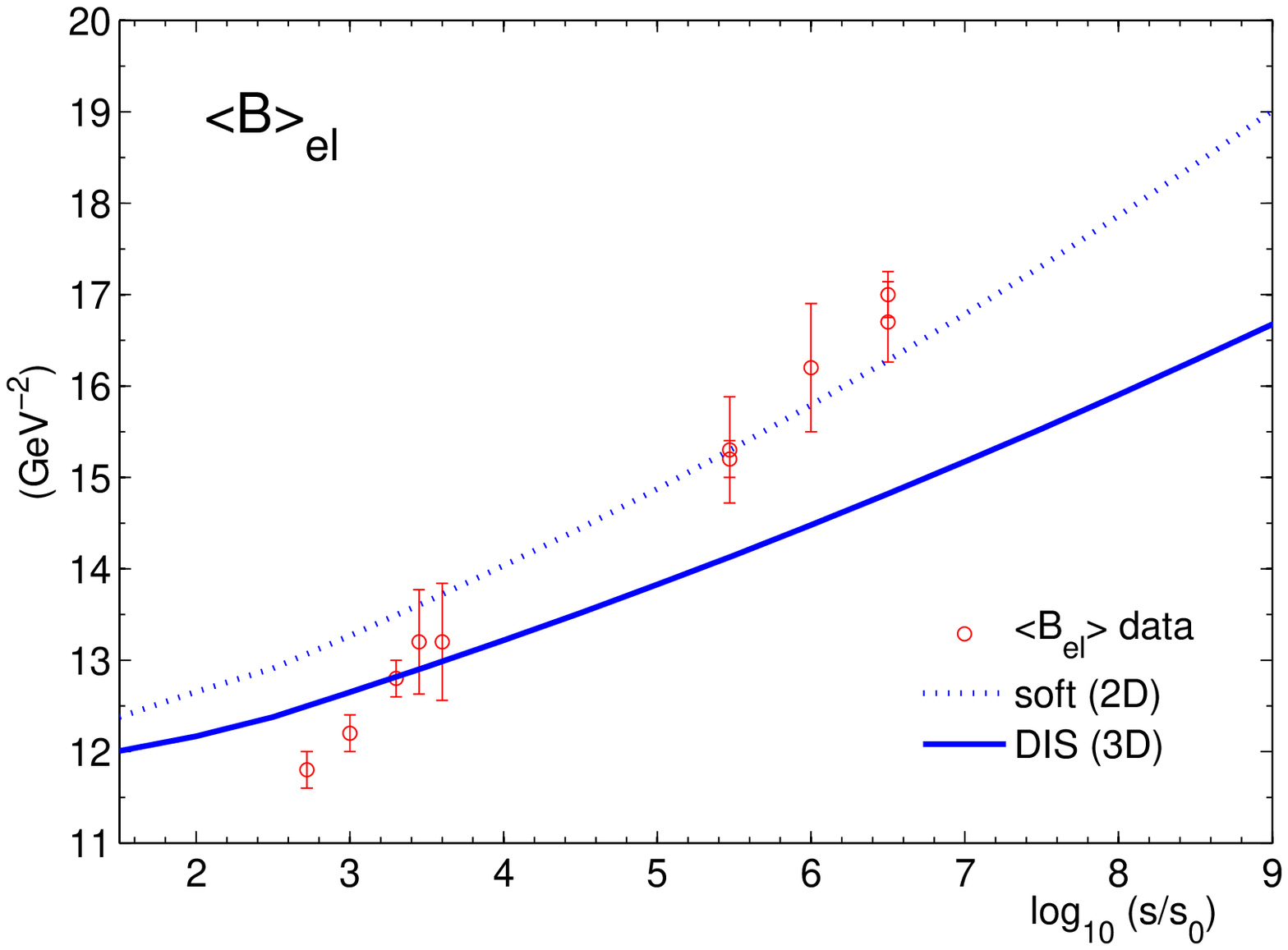}&
\includegraphics[height=.2\textheight]{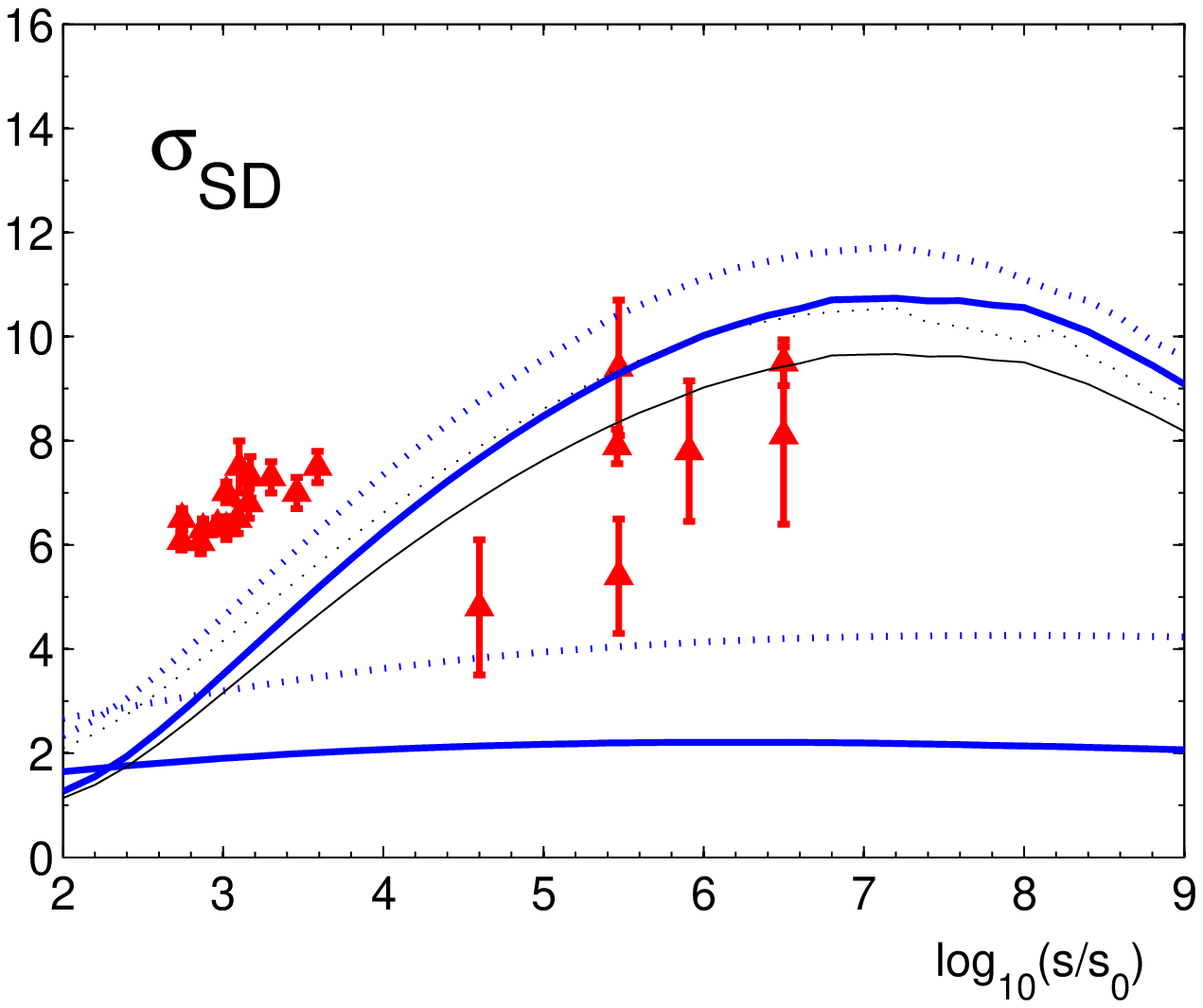}\\
\end{tabular}
\caption{}
\end{figure}

\section{Conclusions}
In the table we show our prediction for the LHC range of energy with comparison with other models.
The most pronounced difference between our estimates and RMK model\cite{RMK} in the value of the survival probability for the central Higgs production. It is worth mentioning, that in our approach the small value of $<|S^2|>$ stems from Good-Walker mechanism but not from the enhanced diagrams as in GLMM model \cite{GLM2}.

We will be happy if you, thinking about Pomeron, would remember that  (i) there is a possibility to describe the data without introducing soft Pomeron; (ii) the idea about the saturation momentum as the only dimensional  parameter  of the  strong interaction works; and (iii) all parameters for strong interaction can be found from DIS.


\begin{theacknowledgments}
We are grateful to Errol Gotsman, Uri Maor and Alex Palatnik for useful  
discussions on the subject.
This research was supported
in part by the Israel Science Foundation, founded by the Israeli Academy of Science
and Humanities, by BSF grant $\#$ 20004019 and by
a grant from Israel Ministry of Science, Culture and Sport and
the Foundation for Basic Research of the Russian Federation.
\end{theacknowledgments}

\end{document}